\documentclass[preprint,superscriptaddress,preprintnumbers,showpacs]{revtex4}

\usepackage{amssymb}
\usepackage{amsmath}
\usepackage{graphicx}
\usepackage{dcolumn}
\usepackage{bm}
\usepackage{multirow}
\begin{document}

\newcommand*{\sjtu}{INPAC, Department of Physics and Shanghai Key Laboratory for Particle Physics and Cosmology, Shanghai Jiao Tong University,  Shanghai, China}\affiliation{\sjtu}
\newcommand*{\ciae}{China Institute of Atomic Energy, Beijing, China}\affiliation{\ciae}

\title{Sterile Neutrino Search Using China Advanced Research Reactor}
\author{Gang Guo}\affiliation{\sjtu}
\author{Fang Han}\affiliation{\sjtu}
\author{Xiangdong Ji}\affiliation{\sjtu}
\author{Jianglai Liu}\email{jianglai.liu@sjtu.edu.cn}\thanks{Corresponding author}\affiliation{\sjtu}
\author{Zhaoxu Xi}\affiliation{\sjtu}
\author{Huanqiao Zhang}\affiliation{\ciae}

\begin{abstract}
We study the feasibility of a sterile neutrino search
at the China Advanced Research Reactor by measuring $\bar {\nu}_e$
survival probability with a baseline of less than 15~m.
Both hydrogen and deuteron have been considered as potential targets.
The sensitivity to sterile-to-regular neutrino mixing
is investigated under the ``3(active)+1(sterile)''
framework. We find that the mixing
parameter $\sin^2(2\theta_{14})$ can be severely constrained by such
measurement if the mass square difference $\Delta m_{14}^2$
is of the order of $\sim$1 eV$^2$.
\end{abstract}

\date{\today $\vphantom{\bigg|_{\bigg|}^|}$}

\maketitle

\section{Introduction}
Neutrino flavor mixing and oscillation, a direct consequence of
non-zero neutrino masses, are well established by experimental
data~\cite{PDG}. Most of the experimental data
to date, including the recent discovery of
$\theta_{13}$ \cite{dayabay,reno,dchooz} at nuclear reactors,
can be described under the 3-flavor mixing framework~\cite{PDG}.

Beyond the standard 3-flavor model, sterile neutrinos are postulated
as a special type of (heavy) neutrinos that do not interact
electromagnetically, weakly, or strongly, hence the name
"sterile". Since 90's in the last century, there have been several
neutrino oscillation experiments
\cite{Gallex,SAGE,LSND,Miniboone}
which seemed to detect anomalies beyond the
3-flavor mixing. Quite recently, after a re-evaluation of the
nuclear reactor flux prediction~\cite{new_flux,new_flux2}, a global deficit
(2-3~$\sigma$ level) is emerging in the measured flux from all short baseline
reactor neutrino experiments~\cite{anomaly}. Global fits under
``3(active)+1(sterile)'' (or ``3+1'' in short) framework favor a
sizable $\sin^22\theta_{14} \sim 0.1$ and a mass
splitting  $\Delta m^2_{14}$ ranging from 
1 eV$^2$ and above (see, e.g. \cite{global_fit, giunti13}).

Although somewhat non-standard, sterile neutrinos appear to be the
simplest explanation of existing experimental anomalies. They are
also candidates for warm or cold dark matter under many theoretical
models~\cite{warm_DM}.

Numbers of experiments are underway worldwide to search for
normal-to-sterile neutrino oscillations~\cite{white_paper}, including several
projects at nuclear reactors. In this paper, we
evaluate the feasibility of carrying
out a short baseline neutrino experiment using the upcoming
state-of-art China Advanced Research Reactor (CARR) reactor~\cite{CARR} with
a thermal power of $\sim60$~MW$_{th}$.

\section{Sterile Neutrino Search at CARR}
As will be demonstrated below, to search for eV-scale sterile
neutrinos via reactor neutrinos (a few MeV in energy) disappearance,
short baseline ($<15$~m) is needed
to have sufficient sensitivity.
Due to safety regulations, it is nearly impossible
to place the detector so close to commercial reactors ($\sim$GW$_{th}$).
In addition, the core size ($\sim$3 meter) introduces smearing effects
to the oscillation signal~\cite{yasuda}. Compact research reactors ($\sim$0.1~GW$_{th}$),
on the other hand, are more advantageous in these regards (with a
cost of lower neutrino flux from the core).

\subsection{CARR experimental site}
China Advanced Research Reactor (CARR), constructed at the China Institute of
Atomic Energy in Beijing, China, is a tank-in-pool, inverse neutron
trap type, light water cooled, heavy water reflected, multi-purpose
research reactor~\cite{CARR}. The reactor body is immersed in a water pool
with 16~m in depth and the core is located 12~m below the pool water
surface. The reactor core is about 0.8~m in height and 0.4~m in diameter.
CARR takes U$_3$Si$_2$-Al as the fuel meat, with a 20\% enrichment
of ${}^{235}$U in weight. With a thermal power of 60~MW, the
maximum output thermal neutron flux is about $1.0\times 10^{15}$n/cm$^{2}$/s.

CARR was designed as a general-purpose thermal neutron facility for
material and biological researches, as well as for isotopic
production/enrichment. At present, there are 9
horizontal and 21 vertical beam lines coupled to user equipments.
On the ground level of the experimental hall (where horizontal beam lines are),
the center of the reactor core is 120 cm above the floor.
The outer diameter of the concrete shielding structure
is 5.0 m. With this geometry, the closest radial location for a
neutrino detector is about 7~m from the core.

\subsection{Neutrino flux and spectrum at CARR}
\label{flux_carr}
Nuclear reactor is a very intense source of neutrinos. Pure
electron antineutrinos $\bar{\nu}_e$s are produced via
$\beta$-decay of fission fragments. For a 1~GW$_{\rm{th}}$ reactor,
there are approximately $2\times 10^{20}$ $\bar{\nu}_e$s emitted per second.
%Neutrinos are
%produced along the $\beta$-decay of fissions fragments, and since these
%unstable fission products are all neutron-rich nuclei, only
%$\beta^{-}$ decays are allowed, the neutrinos are pure $\bar{\nu}_e$s.

Fission nuclei are dominated by 4 isotopes, ${}^{235}$U, ${}^{238}$U,
${}^{239}$Pu, and ${}^{241}$Pu, while other isotopes contribute less than
0.1\%.
%With an average energy of about 200~MeV released and
%six $\bar{\nu}_e$ emitted per fission,
%$2\times 10^{20}\nu/s$ produced for a 1GW thermal power reactor.
Direct theoretical calculation of the neutrino flux and energy spectrum
bare large uncertainties (at the level of 10\%)~\cite{new_flux},
primarily due to incomplete information from nuclear databases.
On the other hand, for ${}^{235}$U, ${}^{239}$Pu, ${}^{241}$Pu, $\bar{\nu}_e$
energy spectra have been derived from measured electron spectra at
ILL, with an average uncertainties less than 2\%, mainly originated
from the uncertainty due to the conversion from electron to neutrino
spectra~\cite{ILL_spec}. For ${}^{238}$U (fast neutron-induced fissions),
only theoretical calculations exist at present~\cite{new_flux, spec_U238}.
The isotopic concentration in fuel evolves with reactor operation time, so
does the fission rate of each isotope. To predict neutrino flux at a given
time, commercial reactor in particular, detailed core simulation
is needed, which carries its own uncertainty.
%A fraction of neutrons generated by ${}^{235}U$ fissions will be
%captured by ${}^{238}U$, leading to the production of isotope
%${}^{239}Pu$ and ${}^{241}Pu$ to a lesser extent.
%During the core burn-up process in one operation cycle,
%${}^{235}U$ depletes and ${}^{239}Pu$ and ${}^{241}Pu$ breed, while
%the ${}^{238}U$ concentration keeps relatively stable.
%Therefore, the fissions rates of each isotopes evolves as a function
%of operation time. To get a more accurate knowledge about them, detail
%core simulation is required. For a commercial power reactors, the
%${}^{235}U$ enrichment is normally around 3-4\%. Fission rates
%from ${}^{235}U$ and ${}^{239}Pu$ dominates, each constituting about
%60\% and 30\% of the total fission rates, when averaged over operation
%time, while the remaining fissions of ${}^{241}Pu$ and fast neutron
%induced fissions of ${}^{238}U$ contribute less than 10\%.
For research reactors like CARR, however, ${}^{235}$U enrichment is much
higher than that of commercial reactors, fissions of which dominate the
total fission rate. For simplicity, we shall ignore burnup effects and
assume a pure ${}^{235}$U neutrino spectrum for CARR in the remainder
of this work.
%To be simple, we will take a pure ${}^{235}U$ neutrino spectrum as input through the whole article, and assume the reactor thermal power stable in 60MW, without considering the effect of ${}^{235}U$ burnup.
The difference in shape between a real and a pure ${}^{235}$U will be considered
as a shape uncertainty (bin-to-bin).

\begin{figure}[!htbp]
\includegraphics[width=3.5 in]{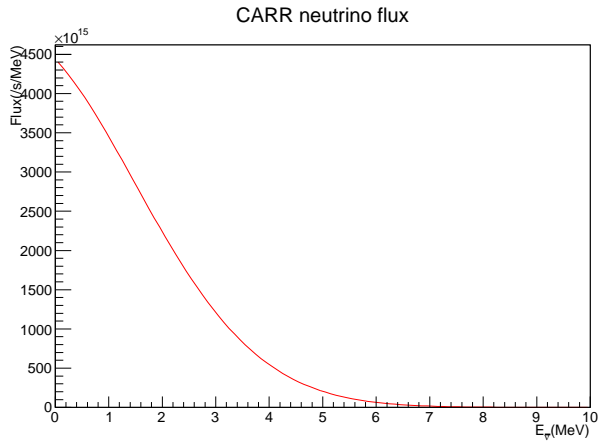}
\caption{Emitted neutrino flux (/s/MeV) at CARR assuming
pure ${}^{235}$U fissions with the parametrization given in \cite{Vogel}.}
\label{fig:flux}
\end{figure}

In this work, we adopted the simple parameterization in \cite{Vogel} for ${}^{235}$U
neutrino energy spectrum,
\begin{eqnarray}
f(E_\nu) = \displaystyle e^{ 0.870 - 0.160 E_\nu -0.0910 E_\nu^2}\,,
\end{eqnarray}
in units of $\bar{\nu}_e/$(MeV $\cdot$ fission),
also shown in Fig.~\ref{fig:flux}.

The energy release per fission for ${}^{235}$U is $201.7\pm 0.6$~MeV
in \cite{ene1} and $201.92\pm 0.46$~MeV in \cite{ene2}.
We take an average with $E_{re} = 201.8$~MeV.
%%%FIXME, check recent Daya Bay publication
For CARR ($P_{th}= 60$~MW), the expected neutrino spectrum
emitted from the core per unit time is:
\begin{eqnarray}
\label{eq:reac_flux}
F(E_\nu) = { P_{th} \over E_{re}} \cdot f(E_\nu)\,.
\end{eqnarray}

\section{ $\bar{\nu}_e$ detection}
Neutrino-target interaction cross section is low, typically of the order of
$10^{-44}$~cm$^2$. Background suppression is a key consideration of such
experiments. On the other hand, short baseline requirement and space
constraint dictate that the neutrino detector has to be placed above
ground without large shielding structure. To supress background,
we investigated three detection techniques, all with timing coincidence
signatures.

\subsection{Inverse-$\beta$ decay with liquid scintillator}
\label{sec:LS}
The classical method for detecting reactor $\bar{\nu}_e$ is the so-called
inverse-$\beta$ decay (IBD),
\begin{eqnarray}
\bar{\nu}_e + p \to e^{+} + n\,.
\end{eqnarray}
Liquid scintillator (LS), usually with $>$10\% hydrogen in mass (11\%
assumed in this study),
is commonly used both as the target and detector
for this reaction. The positron loses kinetic energy immediately and 
annihilates into two 511 keV gammas, emitting prompt scintillation lights.
The neutron will undergo thermalization collisions with hydrogen, and
eventually get captured, emitting gamma rays which are converted into
delayed scintillation lights.
The neutrino energy can be reconstructed via
$E_{e^{+}} \simeq E_\nu - 1.8$~MeV,
where $E_{e^{+}}$ and $E_\nu$ are the kinetic
energy of the positron and neutrino, respectively.
The detected prompt energy $E_p$, on the other hand, contains both the positron
kinetic energy as well as the annihilation energy, i.e.
$E_p = E_{e^{+}} + 1.022$~MeV.
%If we define visible prompt
%energy $E_{vis}$ to include both the kinetic and annihilation energy,
%then
%\begin{eqnarray}
%E_{vis} \simeq  E_\nu - 0.8MeV\,.
%\end{eqnarray}
To enhance neutron detection efficiency and suppress background, most
modern experiments adopt Gadolidium-doped LS (GdLS) $-$ the $\sim$8 MeV
n-Gd capture gamma rays can be used as a clean neutron tag.

\begin{figure}[!htbp]
\includegraphics[width=3.5 in]{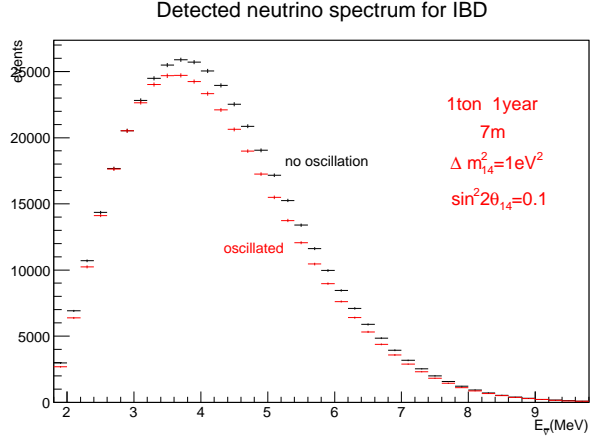}
\caption{Measured IBD neutrino spectrum without (black) and with (red) 
oscillation.
Uncertainties are statistical only.}
\label{fig_ibd_scint}
\end{figure}

Taking into account higher order electroweak corrections,
the cross section of the IBD is given in~\cite{IBD_croX}. Folding it with
reaction neutrino spectrum
(Eqn.~\ref{eq:reac_flux}), detected neutrino spectrum (without oscillation)
can be written as
\begin{equation}
\label{eq:perfect_ibd}
N_{no-osc}(E_{\nu}) = \displaystyle\frac{N_p}{4\pi L^2}\epsilon(E_{\nu})F(E_{\nu})\times\sigma_{IBD}(E_{\nu})\times T\,,
\end{equation}
where $N_p$ is the number of target protons, $\epsilon(E_{\nu})$ is the
detector efficiency, and $T$ is the duration of the measurement.
The resulting neutrino spectrum is shown in Fig.~\ref{fig_ibd_scint}, where
the 1.8 MeV reactor threshold of the IBD is manifest in the curve.
To set the scale, the average detected neutrino rate (assuming 100\%
detection efficiency) is about 7000/MW/ton/year at 7 m.
The oscillated neutrino rate and spectrum will deviate from this spectrum
for a given set of oscillation parameters, as illustrated in
Fig.~\ref{fig_ibd_scint}.

\subsection{IBD with light water}
\label{sec:ibd_with_h20}
A common background for IBD detection with LS is the fast neutron background.
The recoiling protons created by a fast neutron will scintillate and mimic the
prompt energy from the IBD before the neutron get captured.
Since CARR is a surface facility,
fast neutrons background induced by comic muons and from the
reactor itself may pose serious challenge to the experiment.

%Based on our reactor on/off measurement, the intrinsic fast neutron
%background from the reactor could be comparable (FIXME!) to the
%IBD rate.
To mitigate this, if the target is water instead of scintillator,
recoiling protons would not be able to make \v Cerenkov lights,
therefore get rejected. The technology of Gadolinium-doping in water
was proposed in 2004~\cite{Gd_water_doping}
and has been under active development~\cite{Gd_water_RD}, so it would be
possible to maintain
this clean neutron capture tag. However to
use this approach in reactor neutrino experiment there are two obvious 
challenges:
\begin{enumerate}
\item The amount of \v Cerenkov photons is much smaller compared to
the scintillating photons. As a result, water detector has much worse
intrinsic energy resolution compared to that of LS. We would have to
increase the photocathode coverage to get a reasonable amount of
photoelectrons (PEs).
\item The \v Cerenkov threshold for water is
about 289~keV (positron kinetic energy).
The (nonlinear) reconstruction from the visible \v Cerenkov lights to
true neutrino energy will be quite different from the LS case.
%Careful energy calibration with conversion electrons and annihilation gamma
%ray are required.
%Energy calibration becomes a bigger challenge.
\end{enumerate}

\begin{figure}[!htbp]
\includegraphics[width=5 in]{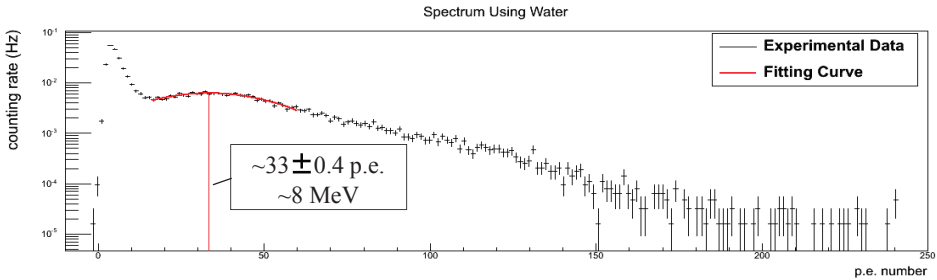}
\caption{Measured muon photoelectron spectrum with a prototype water
detector. The solid angle coverage by photocathode is $\sim$12\%. }
\label{fig:water_light}
\end{figure}

%The first concern can be studied with Monte Carlo.
A realistic estimate of the \v Cerenkov light yield is needed to address
the first concern. Light yields for large water detectors are summarized
in~\cite{PDG} with a range between 3-9 PE/MeV. Small detectors
with less
light attenuation and more reflections could end up with more lights.
To test this, we performed a bench test using
tagged cosmic ray impinging on an acrylic ball with 5 cm diameter
containing pure water. The \v Cerenkov lights are
viewed by four Hamamatsu R7725 photomultipliers close-by. The measured
PE spectrum is shown in Fig.~\ref{fig:water_light}.
The minimum ionization bump ($\sim$8~MeV) is located at $\sim$33 PE.
If we extrapolate this result (photocathode
coverage of $\sim$12\%) to an experiment with $>$50\% coverage,
a light yield of 16 PE/MeV energy deposition would be
attainable. The energy resolution would still be quite low compared to the
LS, but it appears that a 30\% resolution would be achievable at 1 MeV.

The second concern can only be addressed through comprehensive calibration.
Not only one needs a careful calibration with gamma sources, but also
electron sources (beta or conversion electrons) or positron sources to 
establish the energy nonlinearity.
The elaborated calibration programs developed at SuperK~\cite{ref:superk}
and SNO~\cite{ref:sno} provide
invaluable guidance in this regard.

\subsection{$\bar\nu$-D with heavy water}
\label{sec:vD}
Room gamma background, when in accidental coincidence with
random (cosmic or reactor) neutron capture signals, can form IBD-like
background. If such background is significant compared to the IBD signals,
it may become a serious issue (although in principle such a background can be
statistically subtracted).

To further suppress background, we have considered heavy water (D$_2$O)
as a potential target.
In 70's of the last century,
Reines {\it et al.} pioneered $\nu$-D measurements at nuclear
reactors~\cite{reines_vD}.
Reactor $\bar{\nu}_e$s are detected via $\bar{\nu}_e$-deuteron
charge current scattering:
\begin{eqnarray}
\bar{\nu}_e + D \to n+n+e^+
\end{eqnarray}
with the total kinetic energy of the positron given by
\begin{eqnarray}
E_{e^+} \simeq E_\nu - 4 \rm{ MeV}
\end{eqnarray}

The detection signal now becomes a triple coincidence between the prompt
positron signal and two delayed neutron capture signals. Neutron captures
on deuteron will give a single 6.25 MeV gamma ray. To avoid
energy leakage for this high energy gamma, one could dope the heavy water
with salt (NaCl) so that neutron capture on Clorine gives a
total energy of $\sim$ 8.6 MeV, distributed in 2 or 3 gamma
rays~\cite{SNO_salt}. The signals
produced by fast neutron will be singles, the same as in a water detector.
Accidental backgrounds will be highly suppressed by the triple-coincidence
requirement. It should be noted that $\bar{\nu}_e$ can
also scatter off from D via neutral current channel,
$\bar{\nu}_e + D \to n+p+ \bar{\nu}_e$, but with no coincidence signature.

\begin{figure}[!htbp]
\includegraphics[width=3.5 in]{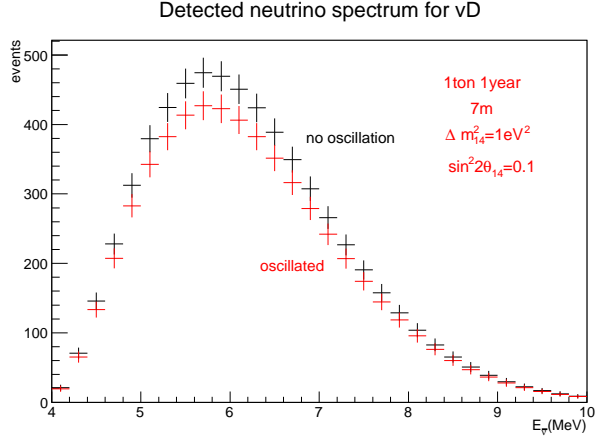}
\caption{Measured $\nu$-D charge current spectrum without (black) 
and with (red) oscillation.
Error bars are statistical only for a 1-ton target at 7 m operating for a
year.}
\label{fig_vd}
\end{figure}

A tabulated charge-current $\bar{\nu}_e$D cross section can be found
in \cite{vD_crox}. The measured spectrum with no oscillations is now
\begin{equation}
\label{eq:perfect_vD}
N_{\rm{no-osc}}(E_{\nu}) =  \displaystyle\frac{N_D}{4\pi L^2}\epsilon(E_{\nu})F(E_{\nu})\times\sigma_{\nu-D}(E_{\nu})\times T\,.
\end{equation}
With a lower cross section and higher
energy threshold,
the total number of detected neutrino events at 7 m is about 92/year/MW/ton
(100\% detection efficiency), 2 orders of magnitude less than that of IBD.
Illustrated in Fig.~\ref{fig_vd} is a comparison of a non-oscillation and
oscillated spectrum.
One should emphasize that the main advantage of D$_2$O is
its unique $\nu$-D charge current signature which may lead to a
background-free measurement.
%In addition, due to higher average
%neutrino energy, this reaction is most sensitive to a different parameter
%space compared to the IBDs (see Sec.~\ref{sec:sensitivity}).

For reference, number of detected neutrino events at CARR using
three different targets (1 ton$\cdot$1 year$\cdot$100\% efficiency)
at 7~m are tabulated in Table~\ref{tab:rate}.
\begin{table}[!htbp]
\renewcommand{\tabcolsep}{0.5cm}
 % \centering
 % \caption{}
  \begin{tabular}{cccc}

    \hline
    Target&Liquid scintillator&H$_2$O&D$_2$O  \\
    Non-osc (/ton/year) & 420129 & 369505 & 5515 \\
    Osc (/ton/year) & 394667 &345329 & 4998 \\
    \hline
  \end{tabular}
  \caption{Event rate at 7 m with 100\% efficiency. Non-osc: $\sin^22\theta_{14}$=0;
    Osc: $\Delta m_{14}^2=1$~eV$^2$, $\sin^22\theta_{14}$=0.1.}
  \label{tab:rate}
\end{table}

\section{Background estimation}
Direct determination of reactor-on background in the CARR experimental hall
has yet to be performed, awaiting for the operation of the reactor.
In this section, background estimation will be given based on current best
knowledge at CARR and projections from other experiments,
serving as a rough guidance to the design of the experiment.

%we prepared a
%LaBr crystal, a $^3$He tube, and a liquid scintillator detector to measure
%gamma, thermal neutron, and fast neutron background directly. Currently
%these measurements await the operation of the reactor.

\subsection{Trigger rate}
Gamma rays (external and internal) 
and cosmic muons are two major contributors to the detector 
raw trigger rate. 

We have not made dedicated gamma spectrum measurement at
CARR, but earlier commissioning run indicated that the dose rate was
less than 3~$\mu$Sv/h, which translate to an upper limit of gamma flux of
150 Hz/cm$^2$, or $<\sim$7.5 MHz for a ton-scale detector. With a 20 cm of Pb
shielding backed up by 5 cm of pure cooper,
the external gamma rate can be cut down to $<$10 Hz ($>$1 MeV
threshold). The internal gamma background can only be suppressed by careful
material screening and selection. For reference, the internal background 
contributes to $<$70 Hz (0.7 MeV threshold) to the trigger rate of Daya Bay detector (20 
ton). If we assume similar materials (stainless steel tank, acrylic 
vessel, etc), similar phototube coverage, and same GdLS in a
ton-scale detector, the internal background would contribute to $<$10 Hz 
for the 1 MeV threshold. 

The integral intensity of vertical muons at the
sea level is about $60$m$^{-2}$s$^{-1}$sr$^{-1}$~\cite{PDG}. The total 
area of the ton-scale detector seen by muons from all directions 
is estimated to be 2 m$^2$, leading to a trigger rate of at least 120 Hz.
Muon-induced backgrounds and possible photomultiplier afterpulsing will also contribute
to the trigger rate. 

Conservatively, we estimate a raw trigger rate of $<$500 Hz, 
which can be comfortably handled by commercial electronics.

\subsection{Time-correlated background rate}
As mentioned above, background in $\bar \nu$-D
charge current channel is hugely suppressed by the triple coincidence
requirement. Here we focus
on the time-correlated background in the IBD channel. We divide the
background into two categories: reactor-associated and reactor-independent.
They can be further sub-divided into correlated and accidental
background.

\subsubsection{reactor-associated background}
\label{sec:reac_asso_bkg}
\begin{figure}[!htbp]
\includegraphics[width=3.5 in]{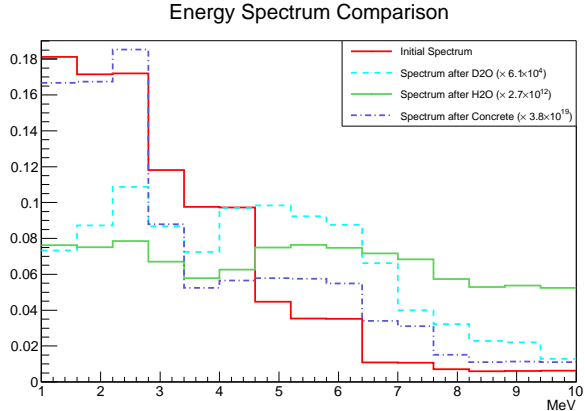}
\caption{Energy spectrum for the residual reactor fast neutron background
predicted by a GEANT4 Monte Carlo.}
\label{fig:nbkg}
\end{figure}

On average, each fission produces 2-3 fission neutrons~\cite{ref:caoj_reactor}.
Thus for a 60~MW$_{th}$ reactor, the total fission neutron flux
from the core is estimate to be $5\times10^{18}$/s.
Most of reactor-associated background is due to neutrons from the reactor
core as well as secondary gammas from neutron capture on metal
or concrete surrounding the core.
%In Table~\ref{}, reactor-on
%fast neutron ($>1$~MeV), slow neutron ($<1$~MeV), and gamma
%rate are summarized. For fast neutrons in the liquid scintillator, if the
%proton recoil energy is above prompt energy threshold, it makes a
%correlated background.
Since CARR is a neutron scattering facility with many neutron guides
from the core, it is difficult to accurately estimate the neutron
background at the detector location. As a start, we assumed an ideal spherical
geometry and used
a GEANT4-based toy Monte Carlo program to transport fission neutrons from the
core through heavy water (1 m), water (1.65 m), and
concrete shielding wall (2.1 m). The neutron spectrum emitting from the
concrete wall is shown in Fig.~\ref{fig:nbkg} together with initial
fission neutrons spectrum.
The shielding factor for neutrons above 1 MeV
is calculated to be $3.8\times10^{19}$ under this geometry.
About one third of these residual
fast neutrons carry a kinetic energy
$>$4 MeV (corresponding to a prompt visible energy $>$ 1 MeV).
So without any neutron shielding around the detector,
and taking into account the detector acceptance ($\sim2\times10^{-3}$
at 7 m), we estimate a correlated rate of $<1\times10^{-4}$/s, two order
of magnitude lower than the IBD signal~\footnote{One should take this
estimate under the caveat that the Monte Carlo had an oversimplified
geometry. The real fast neutron background shall be directly measured.}.
Pulse shape discrimination
is reported to be able to distinguish nuclear recoil from electron
recoil signals in LS~\cite{ref:ps_ls, ref:ps_borexino}.
This could lead to another powerful background suppression factor.

As mentioned earlier, fast neutron background
is absent in the water and heavy water detectors. In all three
detectors, slow neutrons will
only make a single capture signal, contributing to accidental background
only. Our high energy threshold on the delay-like events remove most
gamma background from the natural radioactivity, but not
single neutrons that captured in the target via n-Gd (LS or water)
or n-Cl (heavy water) or high energy gamma rays caused by neutron captured on
envioronmental metal materials (Fe/Cr/Ni etc.)~\cite{ref:nucifier}.
%background only. The accidental background is due to the random coincidence of
%reactor prompt-like gamma rays (1-10 MeV) and single
%neutron-like events (6-12 MeV).
It is difficult to estimate delay-like background without direct
reactor-on measurement. Just for reference, the 
NUCIFIER experiment~\footnote{Very
similar core design as CARR.}
estimated that such a background contribute to B/S of 1:1
for their IBDs~\cite{ref:nucifier} with 10 cm of Pb shielding.
A 20~cm Pb shielding at CARR may further suppress this background.
%Although such a background is statistically subtractable, it is desirable
%that we can cut it down further with more external shielding.

It is interesting to note that other experiments at research reactors also
have measured reactor-correlated background. For example, the ILL reactor
neutrino experiment reported no such background~\cite{ref:ILL}.

\subsubsection{reactor-independent background}
As the detector is placed at the surface, the reactor-independent
background is dominated by neutrons created by cosmic muons (LS detector).
Muons can be separated in two categories: a) ``LS muons'' with long
trajectories in the LS so the detector itself has
100\% tagging efficiecy, b) ``corner muons'' with short or no trajectory
in the LS therefore missed by the detector. Fast neutrons
produced by corner muons will be untagged, producing dangerous background
to the experiment.

To suppress untagged neutron background, we assume a simple outer muon
veto system, e.g. two layers of plastic scintillator paddles, with
an efficiency $95\%$. Let us further assume that the paddles cover an
area of 4 m$^2$ on the surrounding dead materials or corners
which would have been missed by a bare LS detector.
For an area of 4~m$^2$, the muon rate is estimated to be
$60 \times 4 = 240$~Hz.
If we veto all
IBD-like candidate within 200~$\mu$s to a muon detected by the
paddles~\cite{dayabay}, additional deadtime introduced to the experiment is
only 4.8\%.

The residual fast neutron background in the detector
after the paddle veto is estimated using empirical parameterization
in \cite{neutronyield}
\begin{equation}
N_n = 4.14 E_{\mu}^{0.74} \times 10^{-6}/(\mu \cdot \rm{g/cm}^2)\,.
\end{equation}
where $E_{\mu}$ is the muon energy in GeV.
Let us conservatively assume
that the 5\% unvetoed muons each have an effective path length of 100 g/cm$^2$ in
the detector (although they only hit the dead surounding material),
we get an untagged spallation neutron rate of
\[  240 \times 5\% \times 4.14 \times 4^{0.74} \times 10^{-6} \times 100  \sim 0.01 \rm{ Hz} \]
where surface muon average energy $\langle E_\mu\rangle \sim 4$~GeV~\cite{PDG}
has been  been assumed. This background is comparable to the IBD signal rate.
Other background due to muons hitting outside the 4 m$^2$ area is
expected to be small.

Cosmic ray induced fast neutrons in the LS can be further suppressed via
following handles. First, neutron tagging efficiency can be improved by
employing more layers of muon paddles. Second, as illustrated
in \cite{dchooz_prop},
a model independent way to remove all reactor-independent background is to use
the ``reactor on$-$off''. Third, as mentioned in Sec.~\ref{sec:reac_asso_bkg},
pulse-shape discrimation technique may help to
veto neutron recoil signals. For water and heavy water detectors,
on the other hand, there is no correlated background of this nature.

Muon spallation on carbon in the liquid scintillator will produce beta-delayed
neutron emitters $^{9}$Li or $^{8}$He, two classical correlated
background ($\beta$=prompt, n=delayed) for the liquid scintillator-based
experiments. Due to the long half-live (0.18 and 0.12 s), one can not
tag them with parent muons nor is it practical to veto all muons for
a long $\sim$s time window. If we extrapolate from the fitted
results at Daya Bay near site: $\sim$30/day/20-ton for an average
muon energy of 55 GeV and rate of 20 Hz~\cite{dyb_cpc},
we get a surface rate of 1.3/day/ton
assuming that the production yield shares
the same $E_\mu^{0.74}$ dependence as the neutrons. This background is negligible
compared to the IBD rate, and again, can be subtracted by reactor-off data.

%Since the detector is close the the surface, a significant flux of muon-induced fast neutrons is expected, possibly leading to a correlated background, with a rate few times above the neutrino event rate. This background can be efficiently vetoed by muon veto system, with few percent dead time. However, neutrons induced from muon interacting outside the muon veto are only partial rejected. These can be measured during the reactor-off period.
%
%Radioactivity things....
%
%As discussed above, for $\nu$D, it's almost background free due to a triple coincidence signal.

\section{Sensitivity to sterile neutrino search}
\label{sec:sensitivity}
In what follows, we investigate the sensitivity of these neutrino
detector to potential anti-electron to sterile neutrino oscillations. For
simplicity, we shall dwell in the 3(active)+1(sterile) framework.
At short distance($<$15~m) from the reactor, the oscillation from standard
3$\times$3 mixing parameters can be neglected, so the electron anti-neutrino's
survival probability is
\begin{eqnarray}
P_{ee} = 1 - \sin^2 2\theta_{14} \sin^2(1.27 { L\over E_\nu} \Delta m_{14}^2)
\end{eqnarray}
in which $L$ is the baseline (distance from reactor core to detector)
in meter, E is the neutrino energy in MeV, and  $\Delta m^2_{14}$
is the mass square difference between $\nu_4$ and $\nu_1$ mass
eigenstates in eV$^2$. Taking into account the oscillation,
the detected neutrino spectrum for a perfect
detector (Eqn.~\ref{eq:perfect_ibd}, \ref{eq:perfect_vD}) is modified into
\begin{equation}
\label{eq:osc}
N_{osc}(E_{\nu}) = P_{ee}N_{no-osc}(E_{\nu})
\end{equation}

In the reminder of this note, we assume a 60~MW$_{th}$ reactor, 1 year
running time, 1 ton fiducial mass, and 100\% detection efficiency as the
default exposure. Results for different exposure can be projected
straightforwardly.

\subsection{Detector Response Function}
\label{sec:det_response}
As mentioned in Sec.~\ref{sec:LS} and ~\ref{sec:vD},
the positron kinetic energy $E_{e^{+}}$ is simply
related to neutrino energy $E_\nu$.
On the other hand, the detectable prompt energy of the three media is
different. For LS, as mentioned in
Sec.~\ref{sec:LS}, we have $E_p = E_{e^{+}} + 1.022$~MeV. For water and heavy
water detector, due to the \v Cerenkov threshold, the annihilation energy
is hardly visible, so $E_p = E_{e^{+}}$. To get realistic visible energy
spectrum, we convolve $E_p$ with a simple Gaussian smearing (resolution)
and a step-wise threshold function, so
\begin{equation}
\label{eq:visible_energy_spectrum}
N_{vis}(E_{vis}) = \mathcal{T}(E_{vis})\int N(E_{p}) \mathcal{G}(E_{vis}-E_{p}) dE_{p}\,.
\end{equation}
In this expression, $\mathcal{G}(E_{vis}-E_p)$ is Gaussian with a width of
$10\%/\sqrt{E_{p}}$ for liquid scintillator, and $30\%/\sqrt{E_{p}}$ for
water and heavy water (Sec.~\ref{sec:ibd_with_h20}), and
\begin{align}
  \label{eq:threshold}
    \mathcal{T}(E_{vis}) & \left.\begin{array}{l} = 1 \text{    LS}\\ \end{array}\right. \\\nonumber
        &
    \left. \begin{array}{l}
        = 1 (E_{vis}>1 \rm{MeV})\\
        = 0 (E_{vis}<1 \rm{MeV})
    \end{array}\right\} \text{water and heavy water}\,.
\end{align}
%$\mathcal{T}(E_{vis}) =1$ for liquid scintillator, and $=1$ ($E_{vis}>$
%1 MeV) or $=0$($E_{vis}<$1 MeV) for water and heavy water detector.
$N(E_{p})$ is the prompt energy spectrum assuming a perfect detector, i.e.
$N_{no-osc}$ in Eqn.~\ref{eq:perfect_ibd} or \ref{eq:perfect_vD} for no
oscillation hypothesis, or $N_{osc}$ in Eqn.~\ref{eq:osc} when disappearance
is taken into account.

\subsection{$\chi^2$ Definition}
\label{sec:chi2}
The sensitivity of an given experimental setup to a given
set of oscillation parameter ($\sin^2 2\theta_{14}$, $\Delta m^2_{14}$),
in short, is the power that one could differentiate the measured spectrum
from a non-oscillation spectrum. Typically one defines a $\chi^2$ function
as the measure of such difference. In this application, it
should satisfy that 1)  $\chi^2 = 0$ when $\theta_{14} = 0$, 2) for a
given value of $\Delta m^2_{14}$, its variation w.r.t.
$\sin^2 2\theta_{14}$ follows the standard $\chi^2$ distribution. Then to
determine the exclusion limit to a given confidence integral (e.g. 95.5\%),
for each given $\Delta m^2_{14}$ we would scan over the value
of $\sin^2 2\theta_{14}$ to generate measured spectrum, and determine the
boundary of the corresponding $\chi^2$ (e.g. $\chi^2=4$).

Omitting background related systematics~\footnote{
Since we have no direct background measurement nor a shielding
design yet, a particular choice of background would seem improper.
We choose to present background-free
scenario here, and instruct readers to bare this caveat
in mind when reading the sensitivity curves.}, the $\chi^2$ can be
defined as~\cite{ref:chi2}
\begin{eqnarray}
\label{eq_chi2}
\chi^2 =& \displaystyle\sum_{i=1,nbins} {[N_{vis,osc}^i - N_{vis,no-osc}^i \cdot (1+\alpha+\frac{(L+\gamma)^2 }{L^2}+f^i(\eta,\beta))]^2 \over N_{vis,osc}^i[1+\sigma_{b2b}^2 N_{vis,osc}^i]} \\\nonumber
 &\displaystyle{+ ({\alpha \over \sigma_{norm}})^2 +({\eta \over \sigma_{eshift}})^2 + ({\beta \over \sigma_{escale}})^2
+ ({\gamma \over \sigma_{L}})^2 } \,.
\end{eqnarray}
where $N_{vis,osc}^i$ and $N_{vis,no-osc}^i$, respectively, represent the
$i$th energy bin in the visible energy spectrum
(Eqn.~\ref{eq:visible_energy_spectrum}) with and without oscillation.
This is so-called ``rate+shape'' $\chi^2$.
The following systematics have been considered
(see also Table~\ref{tab:sys_pull2}): 1) a 3\% normalization
uncertainty $\sigma_{\rm{norm}}$ (including reactor total neutrino flux,
target protons, and detector efficiency) and its nuisance
parameter $\alpha$;
2) energy non-linearity
including a shift $\sigma_{\rm{eshift}}$ (0.02 MeV) and
a scale factor $\sigma_{\rm{escale}}$ (1\%), and their
corresponding nuisance parameters $\eta$ and $\beta$;
3) 2\% bin-to-bin uncorrelated shape uncertainties $\sigma_{b2b}$, which
is added to the denominator of the first term for simplicity instead of
introducing $N_{bins}$ of pull terms;  $f^i(\eta, \beta)$
represents fractional change of counts in
bin $i$ for a given set of parameter ($\eta, \beta$)
away from (0,0); 4) a
10 cm position accuracy of the center of the core $\sigma_{L}$,
conservatively being assumed to be along the radial direction, and the
corresponding nuisance parameter $\gamma$~\footnote{In principle the
$\gamma$ term can be
absorbed in the normalization uncertainty, but we leave them here
to ease later discussion on two-detector sensitivity.}. The effects of
detector resolution and threshold have been included automatically
by using detected $N_{vis,osc}^i(E_{vis})$ and $N_{vis,non-osc}^i(E_{vis})$
(see Eqn.~\ref{eq:threshold}) in Eqn.~\ref{eq_chi2}.
If one wants to perform a ``rate-only'' analysis, it is equivalent to
using the above $\chi^2$ with a single visible energy bin and set
$\eta$, $\beta$, and $\sigma_{b2b}$ to zero.

The impact of reactor flux can be further suppressed
if we choose to use two identical detectors located at
two different baselines, similar to the setup in the Daya Bay and RENO
experiments~\cite{dayabay,reno}. An earlier independent exploration on 
this approach can be
found in \cite{littlejohn12}. In this case the $\chi^2$ can be redefined as
\begin{eqnarray}
\label{eq_chi2_2det}
\chi^2 =& \displaystyle\sum_{d=n,f;i=1,nbin} {[N_{vis,osc}^{d,i} - N_{vis,no-osc}^{d,i} \cdot (1+\alpha+\epsilon^d+\frac{(L^d+\gamma)^2}{(L^d)^2}+f^i(\eta^d,\beta^d))]^2 \over N_{vis,osc}^{d,i}[1+\sigma_{b2b}^2 N_{vis,osc}^{d,i}]} \\\nonumber
 &+ \displaystyle{({\alpha \over \sigma_{norm}})^2 + ({\gamma \over \sigma_{L}})^2
 + \sum_d[({\epsilon^d\over\sigma_{eff}})^2+({\eta^d \over \sigma_{eshift}})^2 + ({\beta^d \over \sigma_{escale}})^2]}\,,
\end{eqnarray}
in which the superscript $d$ runs between ``near'' and ``far'' to
represent different quantities for the two detectors.
We have also added a detector uncorrelated
efficiency uncertainty $\sigma_{eff}^d$ (0.5\%) and its
corresponding nuisance parameter $\epsilon^d$.
A summary of systematic components, the values, as well as whether
they are correlated between
the two detectors is given in Table~\ref{tab:sys_pull2}.
\begin{table}[!htbp]
 % \centering
 % \caption{}
  \begin{tabular}{|c|c|c|c|}
    \hline
    Systematic uncertainty&Value assumed&Nuisance parameter&Near-Far Correlated  \\
    \hline
    Overall normalization $\sigma_{norm}$ & 3\% & $\alpha$ & C\\
    \hline
    Detector relative efficiency $\sigma_{eff}$ & 0.5\% & $\epsilon^d$ & U\\
    \hline
    Energy shift $\sigma_{eshift}$ & 0.02 MeV & $\eta^d$ & U\\
    \hline
    Energy scale $\sigma_{escale}$ & 1\% & $\beta^d$ & U\\
    \hline
    Reactor spectrum shape $\sigma_{b2b}$ & 2\% & - & U\\
    \hline
    Baseline $\sigma_{L}$ & 10 cm & $\gamma$ & C\\
    \hline
  \end{tabular}
  \caption{Summary of systematic effects included in the two-detector $\chi^2$
    function in Eqn.~\ref{eq_chi2_2det}.
    The last column indicates whether
    the component is correlated between the
    detectors: C=correlated, U=uncorrelated.
    Same systematic uncertainties are assumed for
    single-detector $\chi^2$ (Eqn.~\ref{eq_chi2}) except for
    the detector efficiency, which we have absorbed into the
    normalization uncertainty.
  }
  \label{tab:sys_pull2}
\end{table}

One should note that a couple of conservative
approximations have been made in Eqn.~\ref{eq_chi2_2det}. First, instead
of introducing $nbins$ nuisance parameters for the bin-to-bin shape
uncertainties, we assumed that these uncertainties are also
uncorrelated between the near and far detectors, and lump them to the
denominator of the first term just like the statistical uncertainties.
Second, we have omitted detector correlated energy shift and stretch, and
have assume $\eta^d,\beta^d$ as detector uncorrelated nonlinearity. Both
approximations have been verified to have negligible impact to 
$\theta_{14}$ sensitivity
results.

\subsection{Baseline optimization}
Reactor neutrinos have a energy spectrum
ranged up to 9~MeV, as shown in Fig.~\ref{fig:flux}.
The IBD ($\bar\nu$-D) neutrinos has a energy
peak at $E_{\nu} \simeq 3.7 (5.7)$~MeV. If the true $\Delta m_{14}^2$
is around 1~eV$^2$, as implied by the global analysis, naively one would
put the detector close to the first oscillation maximum to maximize
the analyzing power, i.e.
$L_{osc} = {\pi \over 2} {E_{\nu} \over 1.27 \Delta m_{14}^{2}}$, translating
into 4.6 (7.0) m for IBD and $\bar\nu$-D CC neutrinos.

However, the above discussion is incomplete since we have omitted
influence from
statistics. The fact that the event rate is inversely proportional to $L^{2}$
makes the optimal baseline deviates from the naive
oscillation maximum.

\begin{figure}[!htbp]
\includegraphics[width=3.5 in]{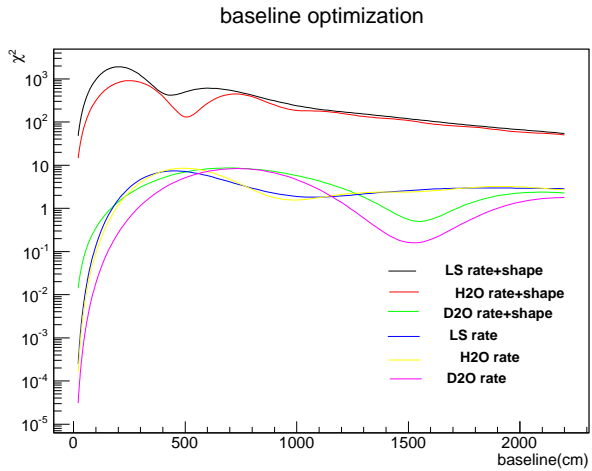}
\caption{Baseline optimization for single detector in ``rate only''
and ``rate+shape'' analyses for LS, light water and heavy water,
with $\Delta m_{14}^2 = 1$~eV$^2$ and $\sin^22\theta_{14} = 0.1$. See text
for details.}
\label{base_1D}
\end{figure}

\begin{figure}[!htbp]
\includegraphics[width=3 in]{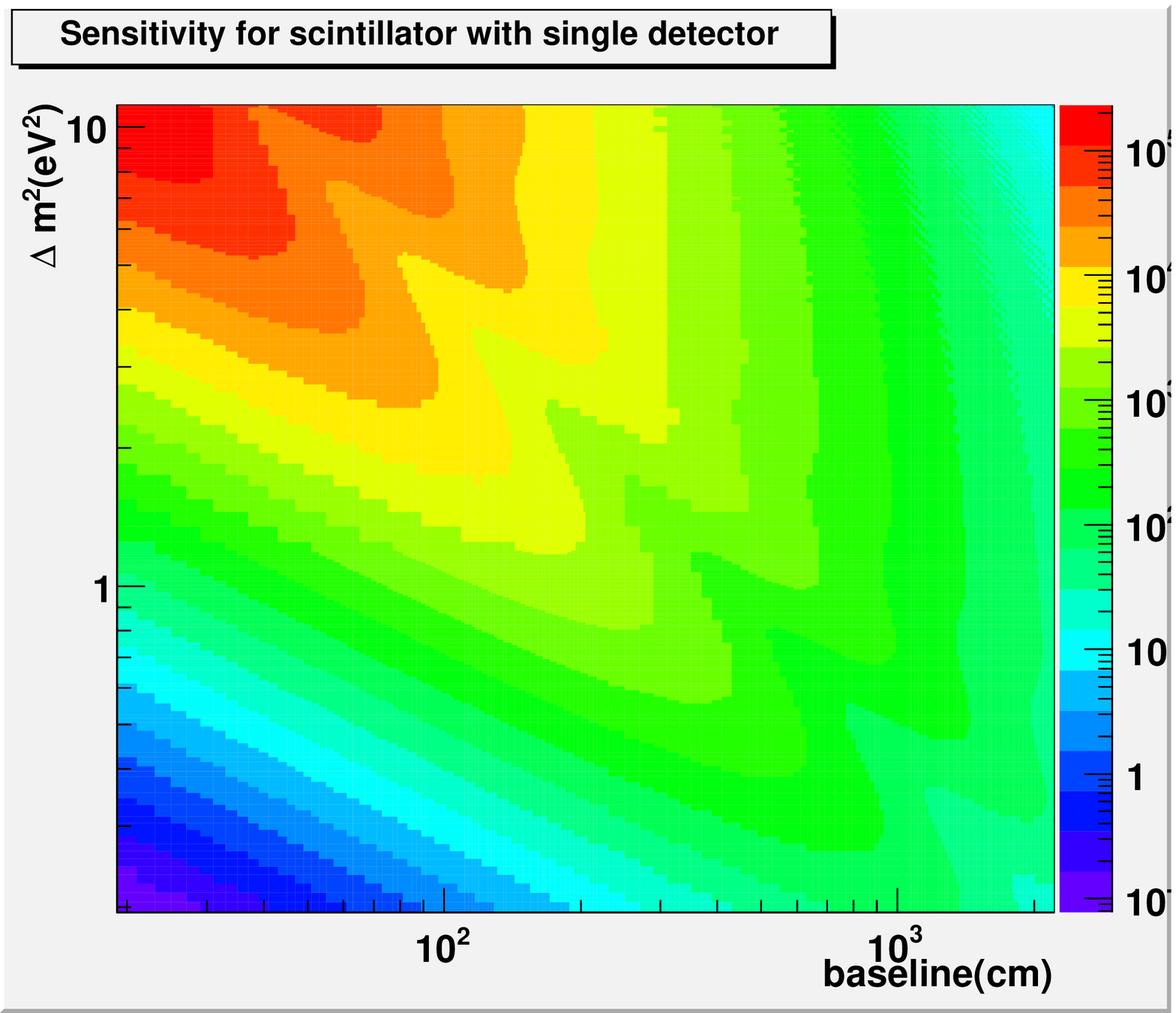}\hspace{0.5 cm}
\includegraphics[width=3 in]{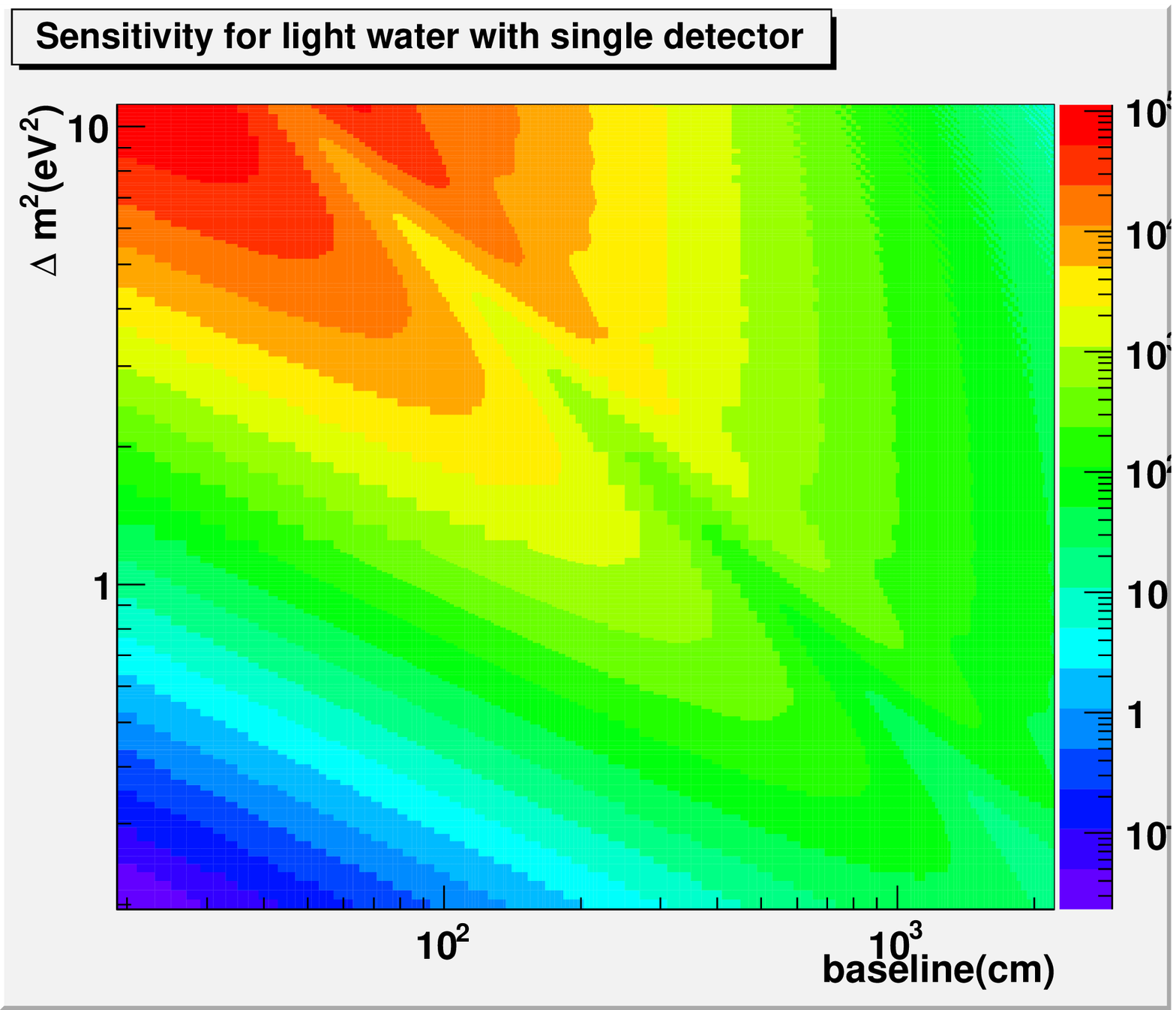}
\hspace{0.5 cm}\includegraphics[width=3 in]{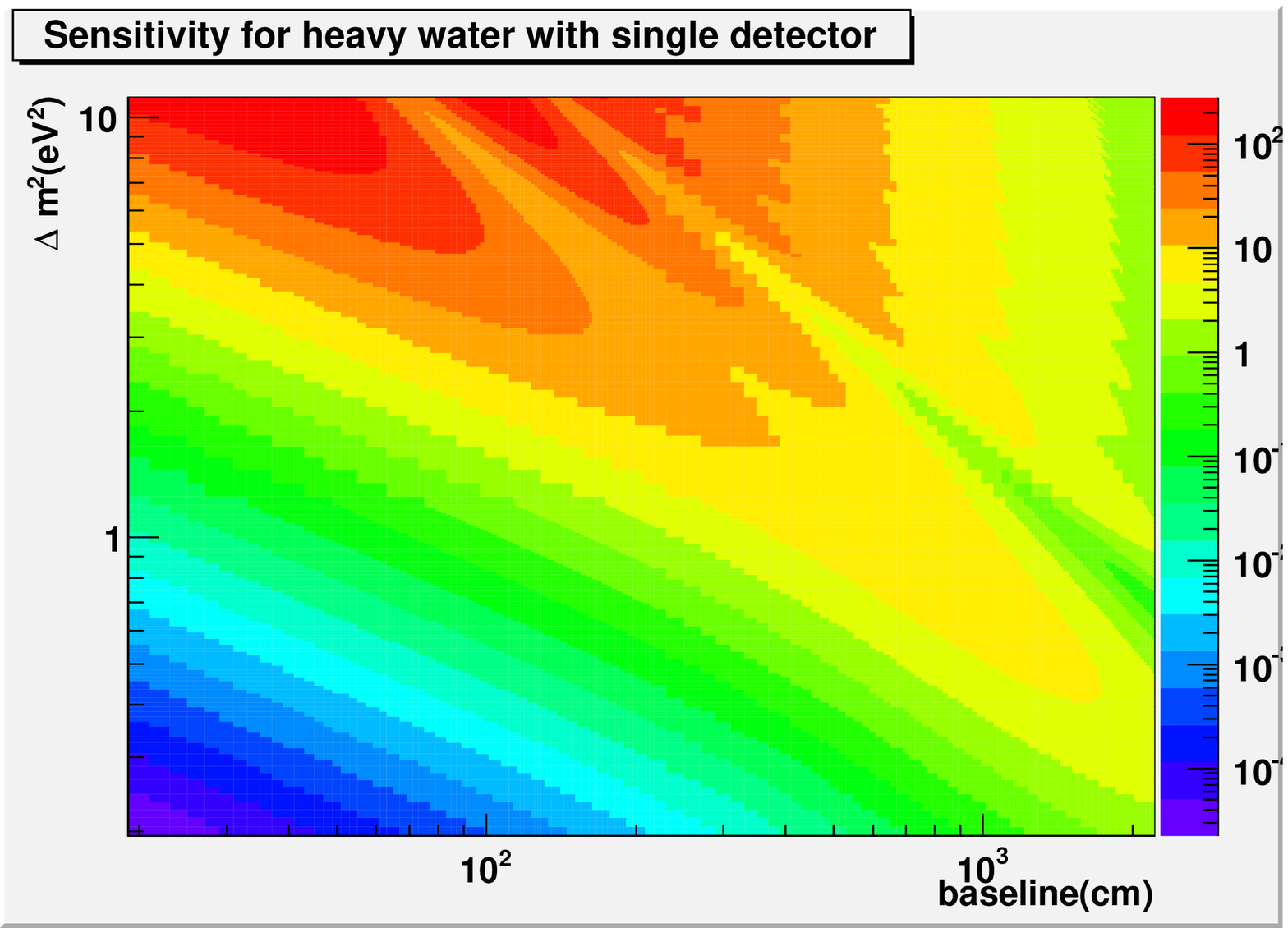}
\caption{$\chi^2$ map vs. baseline and $\Delta m_{14}^2$ for LS
(top left), light water (top right) and heavy water (bottom)
in ``rate+shape'' analysis. 3\% normalization uncertainties has been
considered in the $\chi^2$ calculation,
but not other systematic uncertainties.}
\label{fig:base_2D}
\end{figure}

A more elaborated analysis was made by employing the $\chi^2$ definition
from Sec.~\ref{sec:chi2}. We assumed a fix parameter pair
($\sin^2(2\theta_{14})=0.1$ and $\Delta m_{14}^2 = 1~$eV$^2$), and our
later conclusion does not change significantly with the value of
$\sin^2(2\theta_{14})$. Energy thresholds in Eqn.~\ref{eq:threshold} have
been assumed, but for simplicity we assumed no energy smearing and
set nuisance parameters for energy nonlinearity $(\eta,\beta)$,
baseline uncertainty $(\gamma)$, and the bin-to-bin uncertainty $\sigma_{b2b}$
in Eqn.~\ref{eq_chi2} to zeros. The optimal baseline was determined by
scanning through baseline to find the maximum of $\chi^2$. In
Fig.~\ref{base_1D}, the value of $\chi^2$ vs. baseline
is shown for all three type of detectors with a ``rate-only'' or ``rate+shape''
analysis.
Outside the 5-m shielding wall from the core, coincidentally we observe that 7
meter
%~\footnote{Outside the 5-m shielding wall from the core,
%and with the size of detector shielding structure taken into account,
%the closest distance available of detector placement from the core is
%about 7 m.}
is sufficiently close to the best baseline for all three
detection methods either in ``rate+shape'' or ``rate only''
analysis.  $\chi^2$ map for different values of baseline and
$\Delta m_{14}^2$ have also been shown in Fig.~\ref{fig:base_2D}.
%and
%\ref{fig:base2_2D} for single and two-detector scenarios (``rate+shape'').
One observes a general trend that
the sensitivity for larger value of $\Delta m_{14}^2$
increases with decreasing far detector baseline, which is expected from the
$L\Delta m^2/E$ dependence of the shape.

\begin{figure}[!htbp]
\includegraphics[width=3.5 in]{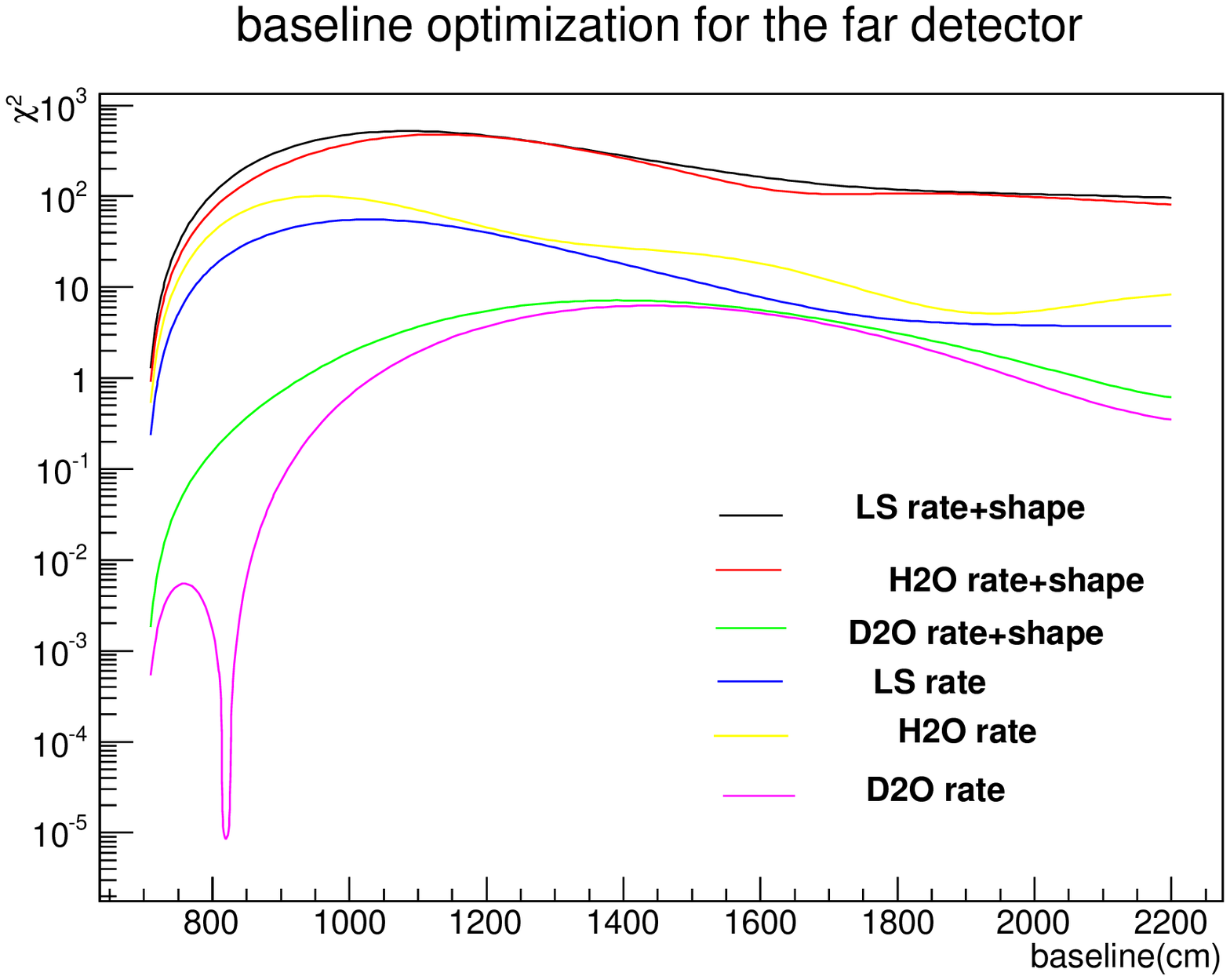}
\caption{Baseline optimization for far detector with the near one
placed at 7 m from the core in ``rate only''
and ``rate+shape'' analyses for LS, light water and heavy water,
with $\Delta m_{14}^2 = 1$~eV$^2$ and $\sin^22\theta_{14} = 0.1$.
See text for details.}
\label{fig:base2_1D}
\end{figure}

For the two-detector scenario, since 7~m is approximately the
closest distance that we can put the detector, we settle the near
detector at this baseline. To determine the optimal baseline for the
far detector, instead of using the two-detector $\chi^2$
in Eqn.~\ref{eq_chi2_2det} (in which one still has to input the expected
``non-osc'' reactor spectrum), we adopted an approach to
construct a pure relative measurement. We used the one-detector $\chi^2$ in
Eqn.~\ref{eq_chi2}, and demanded that
$N_{vis,no-osc}^i = \displaystyle\frac{(L^n)^2}{(L^f)^2}N_{vis,osc}^{n,i}$,
a scaled near
detector spectrum. In general, this approach does not give the best
sensitivity as it entirely omits the theoretical knowledge on reactor
neutrino spectrum. On the other hand, the systematic uncertainty due to
theoretical assumption is also completely avoided. The results of the baseline
scan under this approach are shown in Fig.~\ref{fig:base2_1D}).
The best far detector baseline
is about 11~m for both LS and water in ``rate+shape''
analysis, and about 9~m for LS and 10~m for water in ``rate only''
analysis. The optimal baseline for heavy water is
about 14~m for both analyses. For two-detector discussions in the rest
of this paper, we will assume the optimal baselines in ``rate+shape''
analyses, i.e. a near detector at 7~m, and a far
detector at 11~m (LS \& water) or 14~m (heavy water)~\footnote{
  The final optimization of the far detector baseline will have to
  take into account the actual background levels.
  We also have to make the analysis
  choice of either an absolute measurement using reactor flux calculation,
  or a pure near/far relative non-null search.
  Here we only present a particular choice of the far detector location
  to illustrate the advantage of two-detector scheme.
}.

%FIXME: discussion of rate only sensitivity.

%From the definition,
%$\chi^2 \sim \sin^{2}2\theta_{14} {{1\over L^2} \over { 1\over L}}  \sim   {\sin^{2} 1.27 {L \Delta m^2\over E} \over L}$. To carry on a simple estimation,
%we put $E_0 = 3.7MeV$ in the formulae. From
%$df(L)/dL = 0$, where $f(L) = \alpha \cdot L $, we get $\alpha \cdot L = $.
%
%FIXME: chi2 for two-detectors.
%
%\subsubsection{Single detector}

%\subsubsection{Two identical detector}

%\begin{figure}[!htbp]
%\includegraphics[width=3 in]{base2_s_2D.eps}\hspace{0.5 cm}
%\includegraphics[width=3 in]{base2_w_2D.eps}
%\hspace{0.5 cm}\includegraphics[width=3 in]{base_v_2D.eps}
%\caption{$\chi^2$ map vs. baseline and $\Delta m_{14}^2$ for far detector
%with the near one placed at 7~m from the core, for LS (top left),
%light water (top right) and heavy water (bottom) in
%``rate+shape'' analysis. 3\% normalization uncertainties has been
%considered in $\chi^2$ calculation, but not other systematic uncertainties.}
%\label{fig:base2_2D}
%\end{figure}

\subsection{Sensitivity results}

\subsubsection{Single detector}
\begin{figure}[!htbp]
\includegraphics[width=3.5 in]{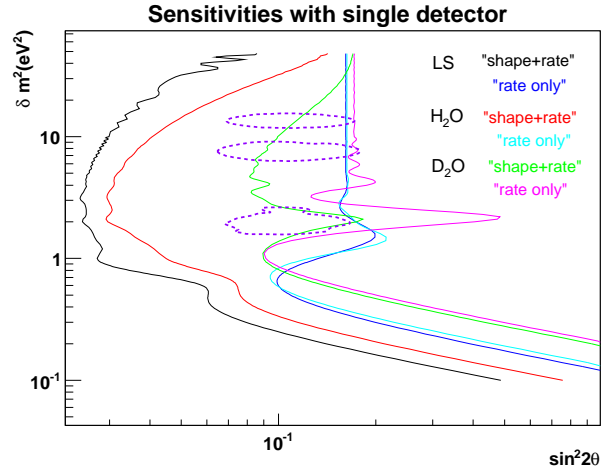}
\caption{
  Sensitivities for a single detector using LS,
  water and heavy water in ``rate+shape'' analysis. All systematics in
  Eqn.~\ref{eq_chi2} have been considered. The dashed contours are the 
  68\% confidence contours in \cite{giunti13} (only those with $\chi^2$ 
  local minima $<1$ are selected to indicate global fits' 
  most favored parameter space and 
  for visual clarity). 
}
\label{fig:sensitivity}
\end{figure}

Using the full $\chi^2$ definition in Eqn.~\ref{eq_chi2},
the 95.5\% exclusion curves for the three detection methods
are shown in Fig.~\ref{fig:sensitivity} in the
$(\sin^22\theta_{14},\Delta m_{14}^2)$ plane.
%Since the sensitivity curves for LS and light water share a lot in common,
%we will discuss them together firstly, and then come to $\nu D$ in a
%very similar way.

With ``rate-only'' analysis, we observe that
for all three techniques, the $2\sigma$ sensitivity limits are
around 0.1 even at optimum $\Delta m_{14}^2$, due to the large
normalization uncertainty. The sensitivity curves for LS and light water
track with each other, although the latter is slightly worse due to a larger
detection threshold. Both experiments are most sensitive to
$\Delta m_{14}^2 \simeq 0.65$~eV$^2$.
For $\Delta m_{14}^2$ larger than  $\sim3$~eV$^2$, the oscillation as a
function of energy becomes so fast that it get smeared out by the energy
resolution of the detector. In this case, one measures a constant
deficit ($\propto 1/2\sin^2(2\theta_{14})$) independent of the baseline, giving
rise to a constant sensitivity at large $\Delta m_{14}^2$. The heavy
water sensitivity is not so much worse than the other two, as the dominating
uncertainty comes from the normalization, not the statistics.
The value of $\Delta m_{14}^2$ where the best sensitivity occurs is
$\sim$1~eV$^2$, higher than that of the IBD, due to higher reaction threshold
($\sim$4~MeV) thereby higher average detected neutrino energy.

The situation is drastically improved with a ``rate+shape'' analysis.
All three exclusion curves moved to much smaller
value of $\sin^22\theta_{14}$ in Fig.~\ref{fig:sensitivity}, LS and water
in particular, as the constraints from the shape will seriously combat
the large normalization uncertainty.
It is also interesting to note that the value of
$\Delta m_{14}^2$ where experiments are most sensitive to has undergone
significant changes compared to that in ``rate-only'' analysis. This
can also be understood as an effect from the shape constraints.
For example, for IBD,
at $\Delta m_{14}^2 \simeq 0.65$~eV$^2$ the overall disappearance
in rate is the largest (on top of the 3\% uncertainty in normalization),
but the shape distortion is
flatter compared to say $\Delta m_{14}^2 = 1$. Therefore it would be relatively
easier to choose a normalization nuisance parameter to balance the spectrum
distortion.
For heavy water, the ``rate+shape'' analysis helps, but not as much as
the IBD, as a result of lower statistics in each energy bin.

\subsubsection {Two detectors}
The sensitivity with two detectors can be investigated in a
similar way using the $\chi^2$ in Eqn.~\ref{eq_chi2_2det}.
Conceptually,
unlike the design of Daya Bay and RENO experiments~\cite{dayabay,reno} (using
well-known $\Delta m^2$), with an unknown $\Delta m_{14}^2$ the ``rate-only''
relative measurement becomes much
dicier. For certain values of $\Delta m^2$,
the normalized event rates at near and far sites would equal to each other
%(Fig.~\ref{fig:rate_ratio}),
therefore would cancel the sensitivity in the
near/far ratio. Under this ``unlucky'' situation, one could still gain
some sensitivity back
by relying on the flux prediction, but the main purpose of
two-detector design would be largely undermined.
%\begin{figure}[!htbp]
%\includegraphics[width=3.5 in]{rate_ratio.eps}
%\caption{ The ratio of normalized total event rate at near
%and far detectors for LS, with no systematic errors considered. We assume
%$\sin^22\theta_{14}=0.1$, $L_n=7$~m, and $L_f=11$~m.}
%\label{fig:rate_ratio}
%\end{figure}

\begin{figure}[!htbp]
\includegraphics[width=3.5 in]{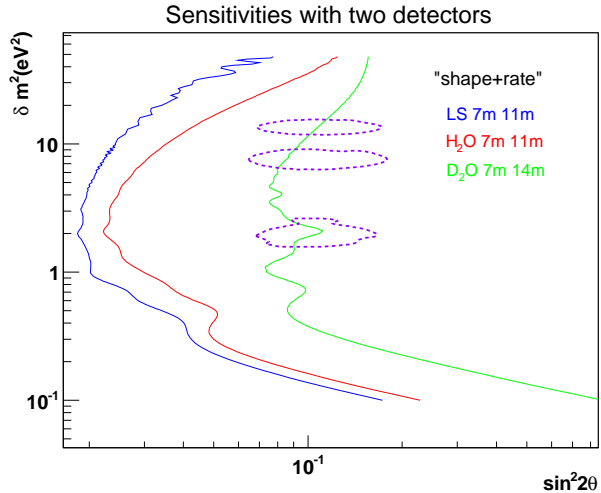}
\caption{Sensitivities for two identical detectors using LS,
  water and heavy water in ``rate+shape'' analysis. All systematics in
  Eqn.~\ref{eq_chi2_2det} have been considered.
  The dashed contours are the 
  68\% confidence contours in \cite{giunti13} (only those with $\chi^2$ 
  local minima $<1$ are selected to indicate global fits' most favored 
  parameter space and 
  for visual clarity).
}
\label{fig:sensitivity2}
\end{figure}

The story is drastically different once the detector shape information is used,
due to the $L/E$ dependence of oscillated spectrum. Within
two-detector scenario, the sensitivity curves with
``rate+shape''
analysis are shown in Fig.~\ref{fig:sensitivity2} for LS, H$_2$O and D$_2$O.
One clearly observes an improved sensitivity
compared to Fig.~\ref{fig:sensitivity} for all detectors. This is anticipated,
since the $\chi^2$ construction in
Eqn.~\ref{eq_chi2_2det} has used information
from both detectors as well as the flux prediction.

%
%
%To understand the shape of the sensitivity curve for $\nu D$ (D$_2$O),
%we can view the reactor neutrinos with mono-energy $E_0$ (4.5MeV for $\nu D$),
%%as we have discussed in Sec. For certain values of $\Delta m^2$, the
%normalized event rate corresponding to neutrinos energy around $E_0$ will be
%equal for the near and far detectors, making the sensitivities worse for
%those $\Delta m^2$.
%These values of $\Delta m^2$ can be calculated via the equation
%$\sin^2(L_n\Delta m^2/E_0) = \sin^2(L_f\Delta m^2/E_0)$, where
%$L_n$ and $L_f$ refer to the baselines for near and far detectors.
%For $\nu D$ (D$_2$O), $L_n$ is fixed at 7~m and $L_f$ is chosen to be 14~m,
%with the optimal sensitivity for $\Delta m^2$ around $1eV^2$.
%
%The above discussion can not simply apply to the IBD (LS and H$_2$O)
%sensitivities with "shape + rate" analysis, since flux uncertainties
%or in a similar way the baseline uncertainties come into play. As in
%single detector, we can fit the normalized event ratio $N^i_f/N^i_n$
%with a global factor 1+$\alpha$, and study the cancelation effect due
%to flux uncertainties and baseline uncertainties.
%
%
%
%
To study contribution from each systematic component, we
compared the variation of the exclusion curve on
($\sin^22\theta_{14},\Delta m_{14}^2$) when
``turning off'' systematics one-by-one. The results for the LS detector
(most sensitive one) at $\Delta m_{14}^2=1$~eV$^2$ is shown in
Table~\ref{tab:sys_uncert_sep}.
Separated systematic uncertainties by taking the quadrature
difference~\footnote{\label{quad} Although quadrature
sum of these effects may not be strictly valid due to correlations,
we take this approach to estimate the relative size of each component.}
are also tabulated in the table.

\begin{table}[htbp]
 % \centering
 % \caption{}
  \begin{tabular}{|c|c|c|}
  \hline
  Errors considered &  Single detector & Two-detector\\
  \hline
   A+B+C+D+E+Stat. & 0.02748  & 0.02017 \\
  \hline
  A+B+C+D+Stat. &  0.02701 &   0.01967      \\
  \hline
  A+B+C+Stat. & 0.02680 & 0.01934    \\
  \hline
  A+B+Stat. & - & 0.01746  \\
  \hline
  A+Stat. & 0.02321 &  0.01691    \\
  \hline
  Stat. only& 0.01055 &  0.00885  \\
  \hline
  \end{tabular}
  \bigskip

  \setlength{\tabcolsep}{8pt}
  \begin{tabular} {ccccccc}
    %\hline
    & Stat & Normalization & Efficiency & E stretch & E shift & Baseline \\\hline
    Single detector & 0.011 & 0.021 & - & 0.013 & 0.004 & 0.005 \\
    Two-detector & 0.009 & 0.014 & 0.004 & 0.008 & 0.004 & 0.004 \\\hline
  \end{tabular}
 %% \\*[0.4ex]
  \caption{Upper table: 2-$\sigma$ sensitivity on $\sin^22\theta_{14}$
    when adding in systematic effects one-by-one.
    Systematic uncertainties include: (A) normalization,
    (B) detection  efficiency, (C) energy
    scale stretch, (D) energy scale shift, and (E)
    baseline uncertainty. The statistical and bin-to-bin
    shape uncertainties are always combined into a ``stat'' uncertainty.
    Target=LS. $\Delta m_{14}^2=1$~eV$^2$.
    Single detector: 7~m. Two-detector: near $@$ 7 m, far $@$ 11 m.
    Lower table: breakdown of uncertainties by taking the quadrature
    differences from the upper table.
    The efficiency uncertainty for a single detector
    has been included into the normalization uncertainty.
  }
\label{tab:sys_uncert_sep}
\end{table}

One sees that two-detector scheme not only improves the statistics, but also
help to reduce the normalization uncertainty. The fact that the normalization
uncertainty does not disappear
completely is a result of the interplay between the
rate and shape constraints. From experimental point of view,
if detectors are constructed
as movable, one could consider a ``swap'' between the near and far
detectors in order to further suppress systematic uncertainties.

\subsection{Effects due to energy and baseline smearing}
The systematic effects discussed above can all be captured in individual 
nuisance parameters $-$ if known to infinite precision, they will not lead to 
biases in neutrino rate or spectrum. Another class of systematics 
introduce smearing to the oscillation signals; one loses sensitivity 
no matter how accurately the smearing is known. Such effects cannot be 
easily incorporated as nuisance parameters in $\chi^2$, therefore require 
separate evaluation. In this section, we discuss two major smearing 
effects: energy 
resolution and baseline smearing. The 
GdLS target is chosen in the study.

\subsubsection{Energy resolution}
\label{sec:energy_resolution}
To extract the impact of energy resolution to $\sin^22\theta_{14}$, we 
recalculated the sensitivity with perfect detector resolution 
(i.e.~no energy smearing), and compare it to that with
realistic energy resolution given in Sec.~\ref{sec:det_response}. 
Simply taking the quadrature difference,
the contribution to $\sin^22\theta_{14}$ sensitivity 
at $\Delta m_{14}^2$ = 1 eV$^2$ is less than $\sim 10^{-3}$ for both 
single and two-detector scenarios.

\subsubsection{Baseline smearing}
\begin{figure}[!htbp]
\includegraphics[width=3.5 in]{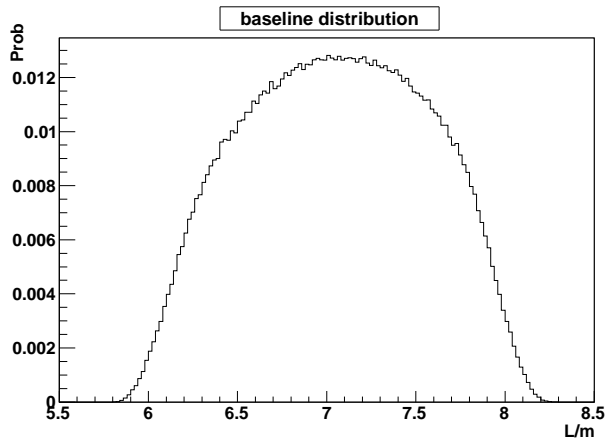}
\caption{Baseline distribution (PDF) for the near detector \@ 7 m. See text
for details.}
\label{fig:L_dis}
\end{figure}

The distance $\bar{\nu_e}$ travels from its origin to IBD interaction point is 
smeared out due to finite-sized core and detector geometry. 
Several control rods are distributed in the core of 
CARR~\cite{ref:carr_diagram}, 
therefore 
neutrino creation points can be approximated as uniform in a 
40~cm diameter and 80~cm height cylinder. In Fig.~\ref{fig:L_dis} 
the baseline distribution for the near detector at 7 m is shown, 
assuming that the target region is a perfect 
cylinder with 1 m diameter and 1.5 m height. 
Obviously such distribution will cause a smearing to the 
L/E oscillation signal. The contribution to $\sin^22\theta_{14}$
sensitivity was evaluated through the same quadrature difference 
procedure above. We obtain a loss of sensitivity of 
less than $2\times 10^{-3}$ in $\sin^22\theta_{14}$ for 
both the single detector and two-detector scenarios for 
$\Delta m_{14}^2$ = 1 eV$^2$.

\section{Discussion and conclusion}
Under the framework of ``3+1'' neutrino mixing,
we have conducted a study of the sensitivities to $\sin^22\theta_{14}$
at short baseline ($<$ 15m) to a
research reactor (CARR) using three targets (LS, H$_2$O, and D$_2$O).
This study suggests that in the absence of background, the LS
detector has the best sensitivity due to higher
IBD reaction rate and
more superior energy resolution. For an experiment detecting
$\bar\nu-$D CC scattering using heavy water, the event rate is far less, with
a powerful suppression of potential background nevertheless.
From the comparison between
``rate-only'' and ``rate+shape'' analyses, we conclude that the spectrum
distortion provide crucial handle to oscillation therefore a key
to the experiment design. To cancel uncertainties from the reactor flux
prediction, we compared the performance of a single-detector and two-detector
design. The latter leads to not only a better sensitivity but also a
suppression of systematic uncertainty. Under the current
best scenario (liquid scintillator, two-detector, no background,
and ``rate+shape'' analysis), a ton-scale
detector operating for a year can reach a sensitivity (95.5\%) of $\sim$0.02
to $\sin^22\theta_{14}$ for $\Delta m_{14}^2\sim$1 eV$^2$, severely constraining
the sterile-to-regular oscillation parameter space suggested by 
global analysis.

\section{Acknowledgments}
This work was done with support from 
the Shuguang Foundation of Shanghai Grant Z1127941, 
and Shanghai Laboratory for Particle Physics and Cosmology at 
the Shanghai Jiao Tong University. The authors acknowledge 
Prof. Karsten Heeger from the Univ. of Wisconsin for an initial 
discussion on sterile neutrinos, and Profs. Weiping Liu and 
Dongfeng Chen from CIAE for 
introducing the CARR reactor to us. We also acknowledge 
Drs. Yufeng Li from IHEP and Bryce R Littlejohn from the Univ. of Cincinnati
for helpful suggestions to the paper draft.

%(1) reactor neutrino flux uncertainties and IBD or $\nu D$ cross sections uncertainties;
%(3) detecting efficiencies errors;
%(4) errors associated with energy calibration;
%(5) systematic errors associated with background,
%the sensitivity calculation above indicate that the setup at a research reactor is sufficient to probe interesting sterile neutrino oscillation.
%
%The missing components in our calculation are a detailed detector design. What's more, we have not consider the smear effect due to the finite size of the reactor core.
%
%It is clear that the background and systematics are driving factors in the design of the experiment. In particular, multiple identical detector modules or segmented detector at different baselines will allow cancelation in reactor flux prediction and detector systematics. As an example, a crude concept of segmented detector is depicted below.


\begin{thebibliography}{0}
\bibitem{PDG}
J.~Beringer {\it et al.} (Particle Data Group), Phys. Rev. D86, 010001 (2012)

\bibitem{dayabay}
F.~P.~An {\it et al.} (Daya Bay Collaboration),
Phys. Rev. Lett. 108 171803 (2012)

\bibitem{reno}
J.~K.~Ahn {\it et al.} (RENO Collaboration),
Phys. Rev. Lett. 108, 191802 (2012).

\bibitem{dchooz} {\it Y. Abe et al.} (Double Chooz Collaboration), Phys. Rev.
Lett. 108, 131801, Phys. Rev. D 86, 052008 (2012)

%\bibitem{t2k} K. Abe {\it et al.} (T2K Collaboration), Phys. Rev. Lett. 107,
%041801 (2011).


\bibitem{Gallex}
P. Anselmann {\it et al.} (GALLEX. Collaboration), Phys. Lett. B 342 (1995) 440;
W. Hampel {\it et al.} (GALLEX Collaboration), Phys. Lett. B 420, 114 (1998);
F. Kaether, W. Hampel, G. Heusser, J. Kiko, and T. Kirsten, Phys. Lett. B 685, 47 (2010).
\bibitem{SAGE}
D. Abdurashitov, V. Gavrin, S. Girin, V. Gorbachev, T. V. Ibragimova, {\it et al.},
Phys. Rev. Lett. 77, 4708 (1996);
J. Abdurashitov {\it et al.} (SAGE Collaboration), Phys.~Rev. C 59, 2246 (1999);
J. Abdurashitov, V. Gavrin, S. Girin, V. Gorbachev, P. Gurkina, {\it et al.},
Phys. Rev. C 73, 045805 (2006); J. Abdurashitov {\it et al.} (SAGE Collaboration),
Phys. Rev. C 80, 015807 (2009).
\bibitem{LSND}
A. Aguilar {\it et al.},(LSND Collaboration), Phys. Rev. D 64, 112007 (2001).
\bibitem{Miniboone}
A. Aguilar-Arevalo {\it et al.} (MiniBooNE Collaboration), Phys.~Rev.~Lett. 102, 101802 (2009);
A. Aguilar-Arevalo {\it et al.} (MiniBooNE Collaboration), Phys.~Rev.~Lett. 105, 181801 (2010).
\bibitem{new_flux}
T. A. Mueller {\it et al.}, Phys. Rev. C 83, 054615 (2011);
\bibitem{new_flux2}
P. Huber, Phys. Rev. C 84, 024617 (2011).
\bibitem{anomaly}
G. Mention, M. Fechner, T. Lasserre, T. Mueller, D. Lhuillier, {\it et al.},
Phys.Rev. D83, 073006 (2011).
\bibitem{global_fit}
C. Giunti and M. Lavede, Phys. Rev. D 84, 093006 (2011).
\bibitem{giunti13} C. Giunti, M. Laveder, Y. F. Li, Q. Y. Liu, H. W. Long, 
Phys.~Rev.~D 86, 113014 (2013)
\bibitem{warm_DM}
See, e.g., T.Asaka and M. Shaposhnikov, Phys. Lett. B 620, 17 (2005).
\bibitem{white_paper}
K. N. Abazajian et al, "Light Sterile Neutrinos: White Paper",  arXiv:1204.5379, April 2012.
\bibitem{CARR}
C.~T.~Ye, China Advanced Research Reactor (CARR): A New Reactor to be Built in China for Neutron Scattering Studies, Physica B 241-243, 48 (1998).

\bibitem{yasuda} O. Yasuda,  JHEP 1109:036 (2011).

\bibitem{ILL_spec}
K. Schreckenbach, G. Colvin and F. von Feilitzsch, Phys. Lett. B160, 325 (1985); F. von Feilitzsch and K. Schreckenbach, Phys. Lett. B118 (1982); A. A. Hahn {\it et al.}, Phys. Lett. B218, 365 (1989).

\bibitem{spec_U238}
P. Vogel, G.K. Schenter, F.M. Mann and R.E. Schenter
Phys. Rev. C 24, 1543 (1981); B. R. Davis, P. Vogel, F.M. Mann and R.E. Schenter,
Phys. Rev. C19, 2259 (1979).

\bibitem{Vogel}
P. Vogel and J. Engel, Phys. Rev. D 39, 3378 (1989).

\bibitem{ene1}
M. F. James, J. Nucl. Energy 23, 517 (1969).
\bibitem{ene2}
V. Kopeikin {\it et al.}, Phys. Atom. Nucl., 67, 1892 (2004).


\bibitem{IBD_croX}
P. Vogel and J. F. Beacom, Phys. Rev. D 60, 053003 (1999).
\bibitem{Gd_water_doping}
J. F. Beacom and M. R. Vagins, Phys. Rev. Lett., 93, 171101 (2004)
\bibitem{Gd_water_RD} Andrew Renshaw {\it et al.}, arXiv:1201.1017


\bibitem{ref:superk}
Super-Kamiokande Collaboration, M.~Nakahata {\it et al.},
Nucl. Instr. Meth. A421 113-129 (1999); E.~Blaufuss {\it et al.}
Nucl. Instr. Meth. A 458, 636-647 (2001); F.~Fukuda {\it et al.},
Nucl. Instrum. Meth. A501, 418-462 (2003).
\bibitem{ref:sno}
SNO Collaboration, J.~Boger {\it et al.}, Nucl. Instrum. Meth. A449 172-207
(2000);
A.~W.~P.~Poon {\it et al.}, Nucl. Instr. Meth. A 452, 15-129 (2000);
M.~R.~Dragowsky {\it et al.}, Nucl. Instr. Meth. A 481, 284-296 (2002);
N.~Tagg {\it et al.}, Nucl. Instr. Meth. A 489, 92-102 (2002);
B.~A.~Moffat {\it et al.}, Nucl. Instr. Meth. A 554, 255-265 (2005);
K.~Boudjemline {\it et al.}, Nucl. Instr. Meth. A 620, 171-181 (2010).

\bibitem{reines_vD}
E. Pasierb, H. S. Gurr, J. Lathrop, F. Reines and H. W. Sobel, Phys. Rev. Lett 43, 96 (1979).
\bibitem{SNO_salt}
  B. Aharmim {\it et al.}, SNO Collaboration, Phys.~Rev.C~72, 055502 (2005).
\bibitem{vD_crox}
  K.~Kubodera and S.~Nozawa, %``Neutrino - nucleus reactions,''
  Int.\ J.\ Mod.\ Phys.\ E {\bf 3}, 101 (1994) [arXiv:nucl-th/9310014].

\bibitem{ref:caoj_reactor} J.~Cao, [arXiv:1101.2266 [hep-ex]]

\bibitem{ref:ps_borexino}
H.~O.~Back {\it et al.} (Borexino Collaboration), Nucl.~Instrum.~Meth.~A584,
98-113 (2008).

\bibitem{ref:ps_ls}
G.~Ranucci, A. ~Goretti, P.~Lombardi, Nucl.~Instrum.~Meth. A412, 374-386 (1998).


\bibitem{ref:nucifier} A. Porta {\it et al.} IEEE TRANSACTIONS ON NUCLEAR SCIENCE, VOL. 57, NO. 5, OCTOBER 2010.

\bibitem{ref:ILL}
H. Kwon {\it et al.}, Phys.~Rev.~D 24, 1097 (1981)

\bibitem{neutronyield}
  Y.~F.~Wang, V.~Balic, G.~Gratta, A.~Fasso, S.~Roesler and A.~Ferrari,
  %``Predicting neutron production from cosmic ray muons,''
  Phys.~Rev.~D~64, 013012 (2001).

\bibitem{dchooz_prop}
F. Ardellier {\it et al.} (Double Chooz Collaboration), hep-ex/0606025 (2006)

\bibitem{dyb_cpc}
  F.~P.~An {\it et al.}  (Daya Bay Collaboration),
  %``Improved Measurement of Electron Antineutrino Disappearance at Daya Bay,''
  Chin. Phys. C 37, 80 (2013).

%\bibitem{chooz}
%  M.~Apollonio {\it et al.}  [CHOOZ Collaboration],
%  %``Limits on neutrino oscillations from the CHOOZ experiment,''
%  Phys.\ Lett.\ B {\bf 466}, 415 (1999)
%  [hep-ex/9907037].

\bibitem{ref:chi2}
See, e.g., P. Huber, M Lindner, T. Schwetz, and W. Winter, Nucl. Phys. B665, 487 (2003)

\bibitem{littlejohn12} K.~M.~Heeger, B.~R.~Littlejohn, H.~P.~Mumm and M.~N.~Tobin, Phys.~Rev~D 87, 073008, 2013

\bibitem{ref:carr_diagram} Wenxi Tian {\it et al.}, Ann. Nucl. Energ., vol. 34, no. 4, pp. 288–296 (2007)

\end{thebibliography}
\end{document}